\pdfoutput=1

\documentclass[letterpaper,english,reprint, aps]{revtex4-1}
\usepackage[T1]{fontenc}
\usepackage{amsmath}
\usepackage{amssymb}
\usepackage{graphicx}
\PassOptionsToPackage{version=3}{mhchem}
\usepackage{mhchem}

\makeatletter

\pdfpageheight\paperheight
\pdfpagewidth\paperwidth

\providecommand{\tabularnewline}{\\}

\usepackage{physics}

\makeatother

\usepackage{babel}
\begin{document}
\preprint{APS/123-QED}
\title{Two-dimensional honeycomb-kagome $\ce{V_{2}O_{3} }$: \\
a robust room-temperature magnetic Chern insulator interfaced with
graphene}
\author{S. Mellaerts\textsuperscript{1}, R. Meng\textsuperscript{1}, M.
Menghini\textsuperscript{1,3}, V. Afanasiev\textsuperscript{1},
J.W. Seo\textsuperscript{2} , M. Houssa\textsuperscript{1,4} and
J.-P. Locquet\textsuperscript{1}}
\affiliation{\textsuperscript{1}Department of Physics and Astronomy, KU Leuven,
Celestijnenlaan 200D, 3001 Leuven, Belgium, \textsuperscript{2}Department
of Materials Engineering, KU Leuven, Kasteelpark Arenberg 44, 3001
Leuven, Belgium, \textsuperscript{3}IMDEA Nanociencia, Calle Faraday
9, E28049, Madrid, Spain, \textsuperscript{4}Imec, Kapeldreef 75,
3001 Leuven, Belgium}
\email{simon.mellaerts@kuleuven.be}

\begin{abstract}
The possibility of dissipationless chiral edge states without the
need of an external magnetic field in the quantum anomalous Hall effect
(QAHE) offers a great potential in electronic/spintronic applications.
The biggest hurdle for the realization of a room-temperature magnetic
Chern insulator is to find a structurally stable material with a sufficiently
large energy gap and Curie temperature that can be easily implemented
in electronic devices. This work based on first-principle methods
shows that a single atomic layer of $\ce{V_{2}O_{3} }$ with honeycomb-kagome
(HK) lattice is structurally stable with a spin-polarized Dirac cone
which gives rise to a room-temperature QAHE by the existence of an
atomic on-site spin-orbit coupling (SOC). Moreover, by a strain and
substrate study, it was found that the quantum anomalous Hall system
is robust against small deformations and can be supported by a graphene
substrate.
\end{abstract}
\maketitle

\section{\label{sec:Intro}Introduction}

To advance electronics at the nanoscale, the exploration and exploitation
of quantum degrees of freedom in materials becomes indispensable.
Two-dimensional (2D) compounds form excellent systems wherein these
quantum degrees of freedom can be exploited, towards electronic and
spintronic applications. A particular group of interest are 2D\emph{
Dirac materials}, which have a Dirac cone with linear dispersion in
their band structure \citep{DiracMaterial}. These materials have
been predicted to exhibit a large variety of exotic properties: massless
fermions \citep{MasslessFermion}, ultrahigh carrier mobility \citep{ultrahighmobility},
(fractional) quantum Hall effects (QHE) \citep{QHE,FQHE}, etc. Furthermore,
there have been great advances in the preparation and growth of 2D
materials in recent years \citep{Novoselov666,VdWMBE,EpitaxyTMD},
which initiated the study of the largely unexplored group of 2D strongly-correlated
Dirac systems (SCDS). As the Dirac cone has shown to be instrumental
for the realization of many non-trivial topological phases \citep{Haldane,KaneMele},
these SCDS uncover a new rich playground where relativistic dispersion,
electron correlations, and topological ordering meet.

\ 

In this work, we perform an \emph{ab initio} study on the 2D single
atomic layer of $\ce{V_{2}O_{3} }$ with HK lattice structure. Firstly,
we confirm the predicted \citep{HKV2O3} ground state properties of
HK $\ce{V_{2}O_{3} }$ and establish that it is an excellent SCDS
candidate with a room-temperature QAHE. The second part of this work
mainly focuses on the experimental feasibility by studying the structure
under compressive and tensile biaxial strain as well as the graphene
supported $\ce{V_{2}O_{3} }$ monolayer system.

\section{\label{sec:Methods}Computational methods}

All spin-polarized calculations were carried out within density-functional
theory (DFT), as implemented in the Vienna \emph{ab initio }simulation
package (VASP) \citep{VASP}. The generalized gradient approximation
(GGA) in the form of Perdew-Burke-Ernzerhof (PBE) \citep{PBE}, projected-augmented-wave
(PAW) potential \citep{PAWmethod}, and the plane-wave basis with
energy cutoff of $550$ eV were used. For the structural relaxation,
a force convergence criterion of $0.005$ eV/$\text{\AA}$ was used
with the Brillouin zone (BZ) sampled by a $12\times12\times1$ $\Gamma$-centered
$k$-point mesh, with a vacuum space of $17\text{\,\AA}$ adopted
along the normal of the atomic plane. To account for the localized
nature of the $3d$ electrons of the $\ce{V}$ cation a Hubbard correction
$U$ is employed within the rotationally invariant approach proposed
by Dudarev et al. \citep{Dudarev}, where $U_{\text{eff}}=U-J$ is
the only meaningful parameter. The self-consistent linear response
calculation introduced by Cococcioni \emph{et al.} \citep{linearresponse}
was adopted to determine $U$. In this way, a Hubbard correction of
$U=3.28$ eV is found (see Supplementary Material), which is close
to the $U_{Bulk}=3$ eV value found in bulk $\ce{V_{2}O_{3} }$ system
\citep{HubbardV2O3}, which is then applied throughout the paper.

The magnetic and electronic self-consistent calculations were performed
with a total energy convergence criterion of $10^{-6}$ eV with the
BZ sampled by a denser $24\times24\times1$ $\Gamma$-centered $k$-point
mesh. To test the sensitivity of the results to the choice of the
functional, the local density approximation (LDA) \citep{LDA} and
the screened exchange hybrid density functional by Heyd-Scuseria-Ernzerhof
(HSE06) \citep{HSE06} are employed. Additionally, for the substrate
calculations, the van der Waals (vdW) interactions were taken into
account by the use of Grimme's DFT-D3 method \citep{Grimmecorrection}.

\begin{figure*}
\includegraphics[scale=0.55]{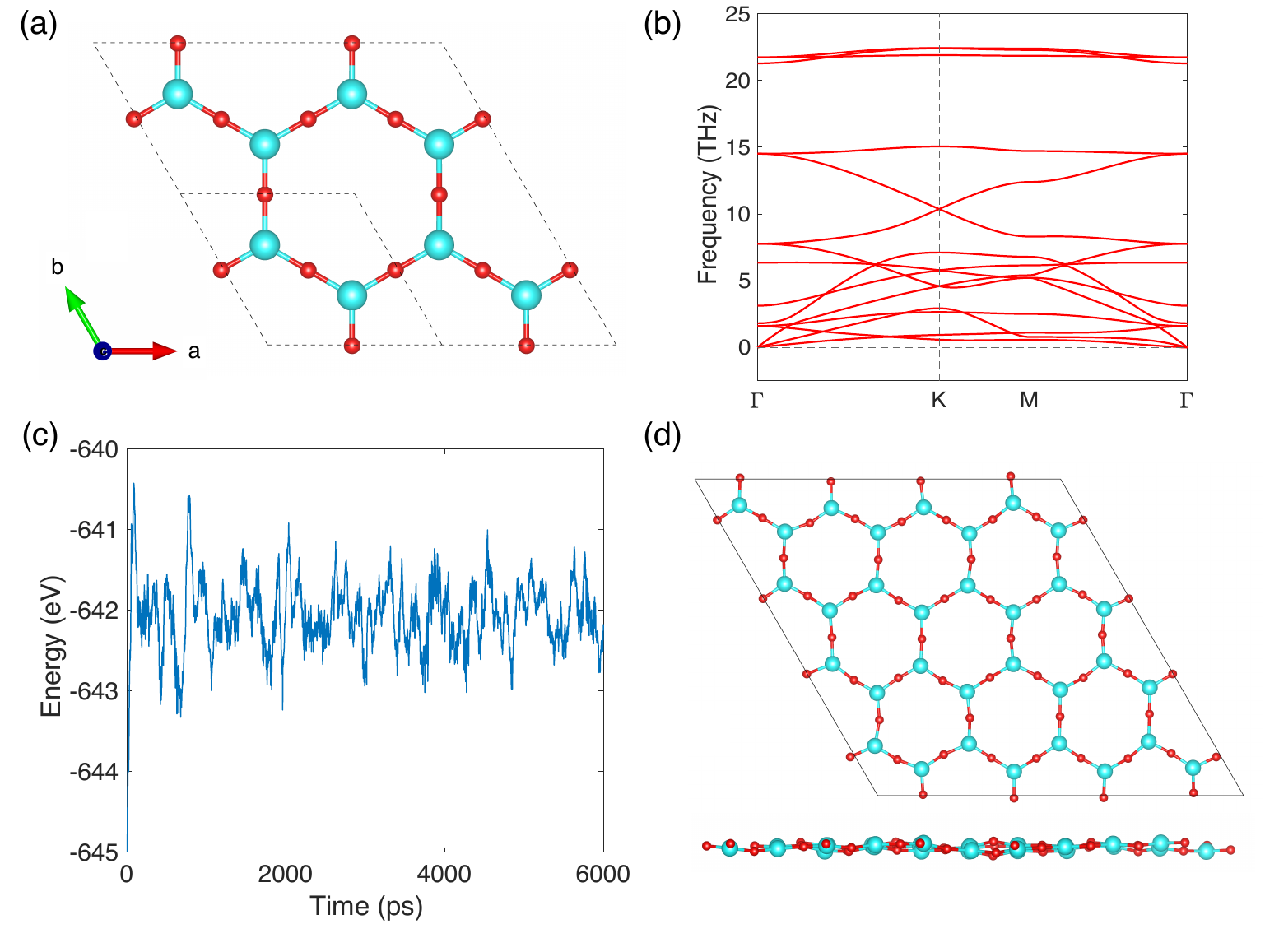}

\caption{$\ce{V_{2}O_{3} }$ monolayer and its structural stability. (a) The
$(2\times2)$ supercell of the HK $\ce{V_{2}O_{3} }$ monolayer with
the unit cell indicated by dashed lines and cyan and red representing
the $\ce{\ce{{V} }}$ and $\ce{\ce{{O} }}$ atoms, respectively. (b)
The calculated phonon band dispersion. (c) The energy fluctuations
during AIMD and (d) the atomic structure obtained after $6000$ ps.}
\label{fig:structural}
\end{figure*}

The phonon dispersion was calculated self-consistently on the basis
of the density functional perturbation theory (DFPT) and with the
use of the PHONOPY package \citep{phonopy}. The \emph{ab initio}
molecular dynamic (AIMD) simulations were carried out on a $4\times4\times1$
supercell at $300$ K in the canonical ensemble using the Nose-Hoover
thermostat approach \citep{NV1,NV2} with $3000$ times steps of step
size $2$ fs. The Curie temperature was estimated by Monte Carlo (MC)
simulations, as implemented in the VAMPIRE package \citep{vampire},
and by the use of a fully quantum-mechanical method where the hamiltonian
is solved by Green's functions \citep{jorenmodel}. For the former
calculations a rectangular $50\times50\sqrt{{3}}$ supercell was used,
where the spins were thermalized for $10^{4}$ equilibrium steps,
followed by $2\cdot10^{4}$ averaging steps to calculate the thermal
averaged magnetization for each temperature. The atomic structures
were visualized by the VESTA program \citep{VESTA}.

\section{\label{sec:Results}Results and discussions}

\subsection{\label{subsec:Structural}Structural stability and mechanical properties}

The structurally optimized unit cell of HK $\ce{V_{2}O_{3}}$ is shown
in Fig. \ref{fig:structural}a with a planar honeycomb-kagome lattice
with lattice constant $6.193\:\text{\AA}$. A reduction of $0.15\:\text{\AA}$
of the $\ce{\ce{{V} }}$-$\ce{\ce{{O}}}$ bond - from $1.94\:\text{\AA}$
to $1.79\,\text{\AA}$ - is found compared to the shortest V-O bond
in bulk $\ce{V_{2}O_{3}}$ \citep{StructuralBulk}. A similar reduction
is observed in other metal oxides \citep{HKAl2O3,HKY2O3,2DTMOscreening}.
This can be attributed to the reduction in coordination number of
the $\ce{\ce{{V}}}$$^{3+}$ cation in this 2D form, where $\ce{\ce{{V}}}$$^{3+}$
has electron configuration $[\ce{Ar}]3d^{2}$ with the two outer $4s$
electrons of vanadium ionically bonded to the oxygen.

To study the dynamical stability of the monolayer system, the phonon
spectrum was evaluated by DFPT. As shown in Fig. \ref{fig:structural}b,
the spectrum shows no imaginary frequency phonon modes, confirming
the dynamical stability. Furthermore, the AIMD simulation at a fixed
temperature of $300$ K also confirmed the thermal stability, since
total energy fluctuations of only $0.3$ \% are observed while planarity
is preserved, as shown in Fig. \ref{fig:structural}c-d.

\begin{table}[h]
\caption{Comparison of the stiffness constants, Young's modulus ($Y_{s}$),
and Poisson's ratio ($\nu$) of different 2D monolayer systems.}
\label{tab:elastic}

\begin{tabular}{|c|c|c|c|c|}
\hline 
 & $C_{11}$ (N/m) & $C_{12}$ (N/m) & $Y_{s}$ (N/m) & $\nu$\tabularnewline
\hline 
\hline 
HK $\ce{V_{2}O_{3}}$ & $73.09$ & $31.84$ & $59.2$ & $0.44$\tabularnewline
\hline 
HK $\ce{Al_{2}O_{3}}$ \citep{HKAl2O3} & $87.75$ & $59.35$ & $47.6$ & $0.68$\tabularnewline
\hline 
Graphene \citep{HKAl2O3} & $353.68$ & $62.20$ & $342.7$ & $0.18$\tabularnewline
\hline 
h-$\ce{BN}$ \citep{HKAl2O3} & $288.26$ & $63.54$ & $274.3$ & $0.22$\tabularnewline
\hline 
\end{tabular}
\end{table}

In addition to the stability verification, the elastic properties
were determined. By symmetry only two independent stiffness constants
$C_{11}$ and $C_{12}$ exist, where we neglect any out-of-plane bending
contribution to the elastic energy density, i.e. the vertical weight
of the stifness constants vanishes. The stiffness constants were determined
by evaluating the total energy of the system under uniform uni- and
bi-axial deformations, over a range of $-10$ \% to $+10$ \%. For
all deformations, the atomic positions were optimized and the corresponding
energies were determined. In this way, the stiffness constants, in-plane
Young's modulus $Y_{s}=(C_{11}^{2}-C_{12}^{2})/C_{11}$ and Poisson's
ratio $\nu=C_{12}/C_{11}$ were calculated. The resulting values are
summarized in Table \ref{tab:elastic}, and compared to other 2D materials.

The mechanical stability could also be confirmed by verifying the
positivity of the strain energy density function $U(\epsilon)=\frac{1}{2}C_{ijkl}\epsilon_{ij}\epsilon_{kl}$
for all $|\epsilon|\le10$ \%. For the 2D hexagonal systems this translates
into satisfying the conditions $C_{11}>0$ and $C_{11}>|C_{12}|$.
Hence, it can be concluded that the single atomic layer of $\ce{V_{2}O_{3}}$
with HK lattice structure is stable on all three levels: dynamical,
thermal and mechanical.

\subsection{\label{subsec:Magnetic}Magnetic ground state}

\begin{figure}[h]
\includegraphics[scale=0.5]{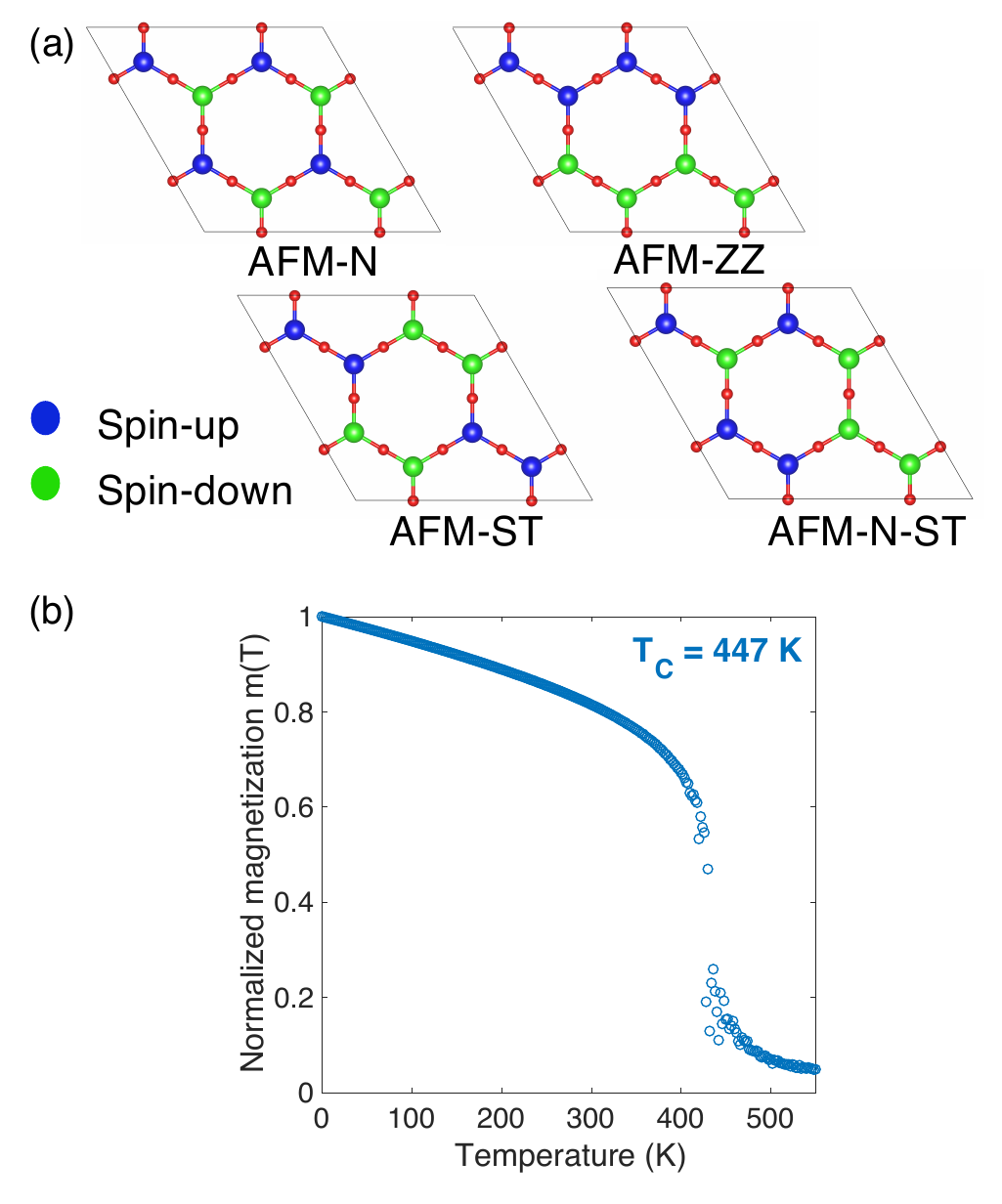}\caption{Magnetic configurations and Curie temperature. (a) Top view of various
possible AFM configurations. (b) The normalized magnetization as a
function of the temperature from the MC simulations indicating a $T_{C}=447$
K. }
\label{fig:magnetic}

\end{figure}

To determine the magnetic ground state of the system various magnetic
configurations have been studied. The total energy of the ion-electron
system was determined for ferromagnetic (FM), paramagnetic (PM), and
four types of antiferromagnetic (AFM) states, depicted in Fig. \ref{fig:magnetic}a.
By comparison of the total energies in Table \ref{tab:magnetism},
it can be concluded that the system is in a FM ground state, with
the AFM stripy (AFM-ST) state as the second lowest energy state. As
a first approximation, the following spin hamiltonian was considered,

\begin{equation}
H=-\sum_{\langle i,j\rangle}J_{1}\mathbf{S}_{i}\cdot\mathbf{S}_{j}-\sum_{i}A(\mathbf{S}_{i}^{z})^{2}\label{eq:SpinHamiltonian}
\end{equation}

where $i$ enumerates the magnetic ions. The first term is the nearest-neighbor
exchange interaction with exchange parameter $J_{1}$, while the second
term corresponds to the single-ion anisotropy with the anisotropy
parameter $A$. The interaction is ferromagnetic when $J_{1}>0$ and
positive $A$ will favor an out-of-plane spin component $\mathbf{S}^{z}$.

\begin{table}[h]
\caption{The total energies (eV/f.u.) for different (out-of-plane) magnetic
configurations with respect to the lowest FM energy state which is
set to zero. }
\label{tab:magnetism}

\begin{tabular}{|c|c|c|c|c|c|c|}
\hline 
\multicolumn{1}{|c|}{$\begin{aligned}\textrm{Magnetic}\\
\textrm{ordering}
\end{aligned}
$} & FM & PM & AFM-N & AFM-ST & \multicolumn{1}{c|}{AFM-ZZ} & AFM-N-ST\tabularnewline
\hline 
\hline 
$\begin{aligned}\textrm{Energy}\\
\textrm{(eV/f.u.)}
\end{aligned}
$ & $0$ & $1.54$ & $0.35$ & $0.18$ & $0.20$ & $0.42$\tabularnewline
\hline 
\end{tabular}
\end{table}

The exchange parameter is evaluated by $J_{1}=[E_{AFM}-E_{FM}]/2zS^{2}$
with the total spin moment $S=1$, and with $z=3$ neighboring spins.
This gives $J_{1}=58$ meV. On the other hand, the magnetocrystalline
anisotropy energy (MAE) defined by $E_{MAE}=E_{[100]}-E_{[001]}$,
with square brackets indicating the orientation of the spins, equals
$2$ meV/f.u. (i.e. $A=1$ meV). This strong MAE can be understood
from degenerate perturbation theory. The SOC-induced interaction involves
the coupling between states with identical spin, with the most important
interaction between the unoccupied and occupied $d$ states. As can
be seen in Fig. \ref{fig:electronic}b, these states at the conduction
band minimum (CBM) at $\Gamma$ and valence band maximum (VBM) at
$\textrm{K}$, respectively consist of the $(d_{xy},d_{x^{2}-y^{2}})$-
and $(d_{xz},d_{yz})$-orbitals. This SOC-induced interaction involves
a change in the orbital angular momentum quantum number $L_{z}$ from
$(d_{xz},d_{yz})$ with $L_{z}=\pm1$ to $(d_{xy},d_{x^{2}-y^{2}})$
with $L_{z}=\pm2$. This interaction between these out-of-plane spins
leads to a maximization of the SOC energy stabilization, as described
in \citep{SOC_anisotropy}.

To obtain an estimate of the Curie temperature ($T_{C}$), the Curie-Bloch
equation in the classical limit,

\begin{equation}
m(T)=\left(1-\frac{T}{T_{C}}\right)^{\beta}
\end{equation}

with $T$ the temperature and $\beta$ a critical exponent, is fitted
to the normalized magnetization obtained by the MC simulation (see
Fig. \ref{fig:magnetic}b). The resulting $T_{C}$ is equal to $447$
K, and can also be compared within the mean field approximation by
using the formula

\begin{equation}
T_{C}=\frac{{S(S+1)}}{3k_{B}}J_{1}
\end{equation}

with Boltzmann constant $k_{B}$ , which gives $449$ K. This is in
relative close agreement with the MC result, which can be attributed
to the relatively high MAE.

\ 

These MC simulations treat the spins as classical vector quantities.
Alternatively, the semi-classical Holstein-Primakoff approximation
is often employed. However, both approximations remain only justified
for high spin values and very low temperatures. Therefore, a fully
quantum mechanical method is employed, where the quantum Heisenberg
hamiltonian is solved by Green's functions \citep{jorenmodel}. Since
the N\'{e}el AFM phase is not the second lowest energy state, it can
be expected that second- and third-neighbor exchange interactions
($J_{2},J_{3}$) are playing a dominant role in the magnetic ordering.
Therefore, the DFT total energies were mapped onto the Ising hamiltonian
to obtain $J_{1}$, $J_{2}$ and $J_{3}$ by the following four equations:

\begin{equation}
E_{FM/AFM-N}=E_{PM}-(\pm3J_{1}+6J_{2}\pm3J_{3})S^{2},
\end{equation}

and

\begin{equation}
E_{AFM-ZZ/ST}=E_{PM}-(\pm J_{1}-2J_{2}\mp3J_{3})S^{2}.
\end{equation}

The $J_{1}$, $J_{2}$ and $J_{3}$ values were found to be $41.9$
meV, $-3.8$ meV and $17.4$ meV, respectively. Taking into account
these second- and third-neighbor exchange interactions in the spin
hamiltonian (\ref{eq:SpinHamiltonian}), the hamiltonian is solved
by the Green's functions, giving a Curie temperature equal to $488.46$
K.

\subsection{\label{subsec:Electronic}Electronic structure}

\begin{figure*}
\includegraphics[scale=0.66]{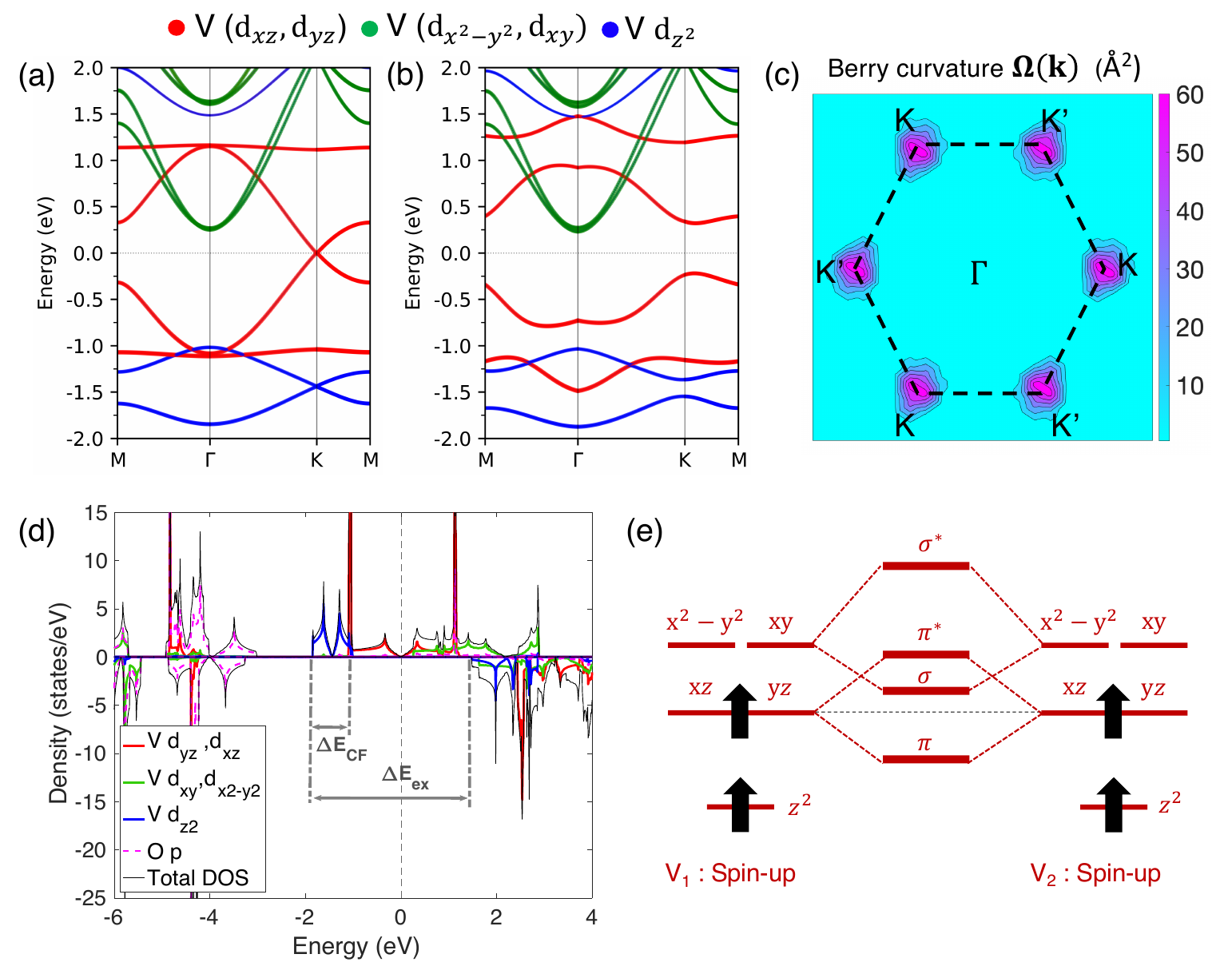}\caption{Electronic band structure with derived orbital and topological character.
(a) The electronic band structure within GGA+$U$ without SOC and
(b) with SOC. (c) The calculated Berry curvature projected onto the
BZ. (d) The orbital- and spin-resolved DOS. (e) The derived orbital
ordering and bonding of the spin-polarized $\ce{\ce{{V}}}$ $3d$
orbitals.}
\label{fig:electronic}
\end{figure*}

The electronic band structure and density of states (DOS) within GGA+$U$
are depicted in Fig. \ref{fig:electronic}. The electronic band structure
shows two key features: a Dirac cone at the \emph{$\mathrm{K}$} point
and the existence of two flat-bands at $\pm1$ eV forming a degeneracy
with the dispersive bands at the $\Gamma$ point. From the DOS it
becomes clear that there exists a strong spin-polarization with the
two characteristic band features being derived from the spin-up $\ce{V}$
$(d_{xz},d_{yz})$-orbitals. This means that the Dirac cone is completely
spin-polarized, making it a Dirac spin-gapless semiconductor (DSGS)
\citep{WangSGS}. By taking into account the intrinsic SOC, it is
found that an energy gap is opened at the $\mathrm{K}$ point, resulting
in an indirect energy gap of $0.45$ eV. It turns out that the energy
gap is strongly dependent on the Hubbard correction $U$ (see Supplementary
Material), which has led to the idea of the cooperative effect between
electron correlations and SOC. The relatively large energy gap opened
at the Dirac point can be understood from the fact that the orbital
degeneracy of the occupied states allows an \emph{atomic} on-site
SOC, involving no hopping process \citep{atomicOnsiteSOC}. This is
in contrast to the second-order hopping SOC in the Kane-Mele model,
or first-order hopping SOC in the Xenes \citep{Xenes}. Within this
understanding, it is clear that an increase of $U$ will force the
electrons into a more localized state implying an increased effect
of the atomic SOC \citep{OwnPaper,MellaertsSimon2020TEi2}.

In addition to the GGA+$U$, the LDA and hybrid exchange-correlation
functional HSE06 were used to study the electronic properties of the
system. The resulting energy gap values are summarized in Table \ref{tab:energygap},
and the calculated band structures can be found in the Supplementary
Material. It is noted that there is a strong variation of the energy
gap depending on the chosen functional. This can be expected as these
functionals are approximating the localized nature of the $3d$ orbitals
very differently, which will be crucial for the atomic SOC effect.

\begin{table}[h]
\caption{The calculated direct and indirect energy gap $E_{g}$ values for
the different exchange-correlation functionals with SOC included.}
\label{tab:energygap}

\begin{tabular}{|c|c|c|c|}
\hline 
 & LDA+$U$ & GGA+$U$ & HSE06\tabularnewline
\hline 
\hline 
Direct $E_{g}$ (eV) & $0.37$ & $0.54$ & $1.54$\tabularnewline
\hline 
Indirect $E_{g}$ (eV) & $0.19$ & $0.45$ & $1.52$\tabularnewline
\hline 
\end{tabular}

\end{table}

From the DOS it can be inferred that the crystal field splits the
$d$-orbitals into three orbital levels whose energy ordering can
be derived from the orbital orientation. The $d_{z^{2}}$-orbital
level has negligible overlap with the $\ce{\ce{{O}}}$ $2p$ orbitals
and thus has the lowest energy state, the ($d_{xz},d_{yz})$-orbitals
oriented out-of-plane will form a $\pi$-bond with the oxygen $\ce{\ce{{O}}}$
$p_{z}$ as bridging ligand, and the $(d_{xy},d_{x^{2}-y^{2}})$-orbitals
will form an in-plane $\sigma$-bond with bridging orbitals $\ce{\ce{{O}}}$
$(p_{x},p_{y})$. This latter bond involves a strong orbital overlap
and therefore the involved orbitals will be energetically less favorable,
forming the highest energy state. The resulting orbital ordering is
shown in Fig. \ref{fig:electronic}e. It should be emphasized that
in the vicinity of the Fermi level, there is approximately no $\ce{\ce{{O}}}$
$2p$ orbital weight, which confirms the localized nature of the two
$\ce{V}$ $3d$ electrons. Based on Griffith's crystal field theory
\citep{griffith1957ligand}, the spin state of the $\ce{\ce{{V}}}$$^{3+}$
cation can be confirmed. By the relative strengths of the exchange
($\Delta E_{\textrm{ex}}$) and crystal field splitting ($\Delta E_{\textrm{CF}}$)
with derived values of $3.34$ eV and $0.8$ eV - it can be concluded
that the $\ce{\ce{{V}}}$ cation has a high ($S=1$) spin state with
magnetic moment $2\mu_{B}$.

\ 

To gain further insight in the nature of the chemical bond, the atomic
charges were determined by the Bader charge analysis code \citep{Bader}.
It is found that the atomic charges on $\ce{\ce{{V}}}$ and $\ce{\ce{{O}}}$
are resp. $Q_{V}=1.59e$ and $Q_{O}=-1.07e$, confirming the ionic
bonding character where the electrons of the $\ce{\ce{{V}}}$ cation
are attracted towards the $\ce{\ce{{O}}}$ anion. Nevertheless, the
atomic charges are slightly lower than in the bulk parent, which suggests
that there is an increased electron delocalization and thus stronger
bonding covalency compared to the bulk. This behavior can be linked
to the reduced $\ce{\ce{{V}}}$-$\ce{\ce{{O}}}$ bond length. On the
other hand, the relative strong ionicity of the bond might explain
the energetic stability of the planar configuration of these TMO monolayer
systems, since the buckling of the $\ce{\ce{{V}}}$-$\ce{\ce{{O}}}$
bond would result in an energetically unfavorable large dipole moment
normal to the atomic plane. In this way, the reduced bond length and
enhanced covalency can be explained as a means to suppress the possible
large dipole moment. These trends and their explanations were already
pointed out for other TMOs \citep{HKAl2O3,HKY2O3}.

\ 

To study the topology of the band structure the Berry curvature $\boldsymbol{\Omega}(\mathbf{k})$
of the system was calculated directly from the DFT calculated wave
functions by the VASPBERRY code, which is based on Fukui's method
\citep{VASPBERRY}. The resulting Berry curvature, shown in Fig. \ref{fig:electronic}c,
becomes non-zero at the Dirac points $\mathrm{K}$ and $\mathrm{K'}$.
By integrating the Berry curvature $\boldsymbol{\Omega}_{n}(\mathbf{k})$
of each $n$-th energy band over the whole BZ and summing over all
occupied bands, the Chern number

\begin{equation}
C=\sum_{n}\frac{1}{2\pi}\int_{BZ}d^{2}\mathbf{k}\,\boldsymbol{\Omega}_{n}(\mathbf{k})
\end{equation}

is found to be $C=1$. Therefore, it can be concluded that the system
is a room-temperature Chern insulator.

\begin{figure*}
\includegraphics[scale=0.62]{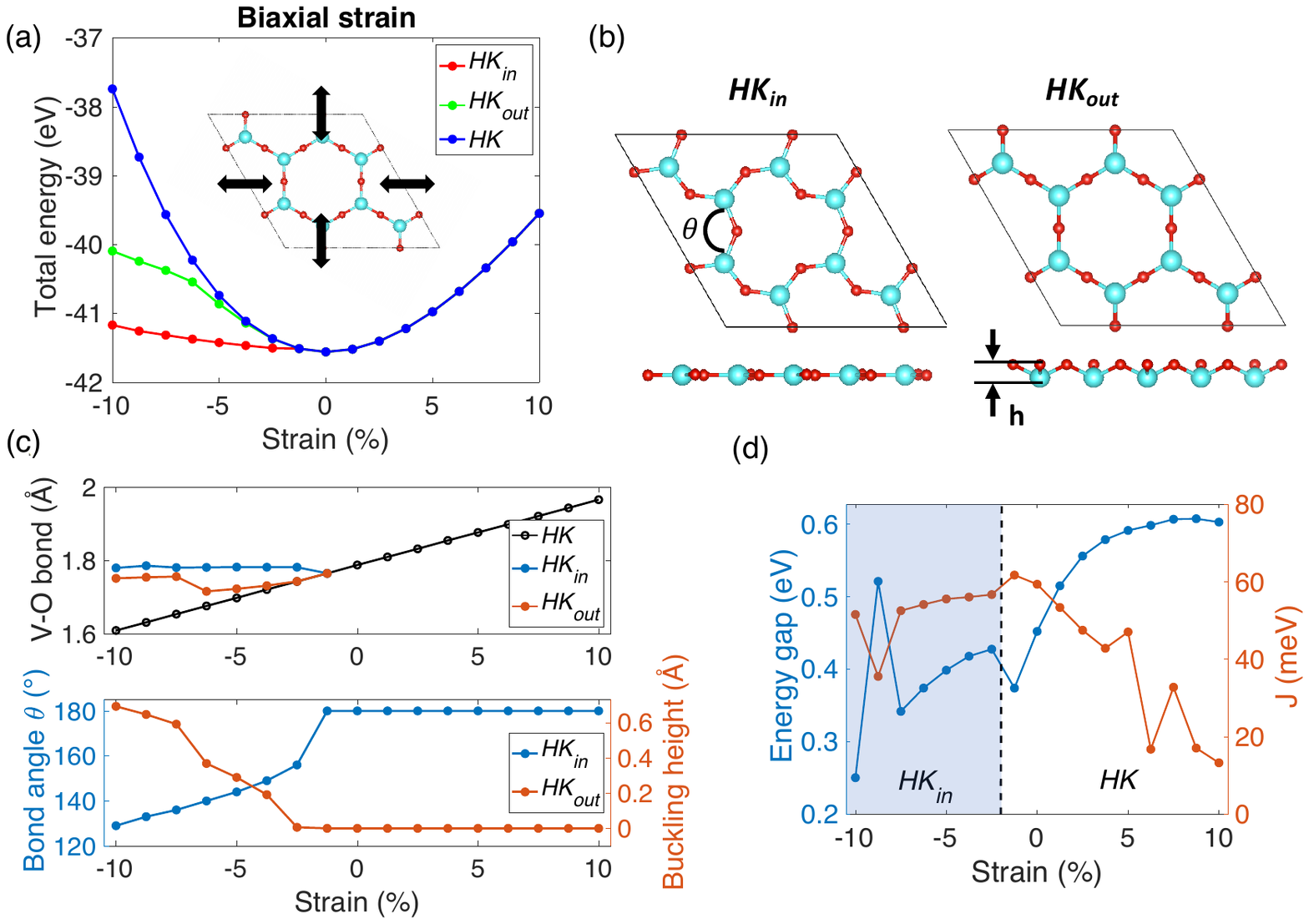}\caption{Biaxial strain study. (a) The total energy of the system under biaxial
strain, taking into account (b) the possible structural deformations
($HK_{in}$, $HK_{out}$) with deformation parameters the bond angle
$\theta$ and the buckling height $h$ respectively. (c) Top: the
$\ce{\ce{{V}}}$-$\ce{\ce{{O}}}$ bond length as a function of strain
for all three configurations. Bottom: the bond angle $\theta$ and
and buckling height $h$ in function of the strain for resp. $HK_{in}$
and $HK_{out}$. (d) The energy gap and magnetic exchange interaction
as a function of strain.}
\label{fig:biaxial}
\end{figure*}

\subsection{\label{subsec:Biaxial}Biaxial strain }

TMOs are known to exhibit a rich phase diagram as a function of strain,
in particular, for bulk $\ce{V_{2}O_{3}}$ the room-temperature metal-insulator
transition (MIT) between the paramagnetic insulating and metallic
state can be realized by the application of epitaxial strain \citep{PiaThesis}.
Therefore, a biaxial strain study on the 2D atomic layer of $\ce{V_{2}O_{3}}$
was performed. The atomic positions were optimized for all biaxial
strained configurations from $-10$ \% to $+10$ \%. It was found
that under compressive strain, the lattice structure undergoes an
in-plane buckling of the $\ce{\ce{{V}}}$-$\ce{\ce{{O}}}$ bond ($HK_{in}$),
as can be seen Fig. \ref{fig:biaxial}a-b. This structural distortion
is found by ionic relaxation with high force convergence criterion
of $0.005$ eV/$\text{\AA}$. On the other hand, by lowering the force
convergence criterion to $0.05$ eV/$\text{\AA}$, two metastable
structures were found in the compressive region; one preserving the
HK lattice structure ($HK$), while the other undergoes an out-of-plane
buckling of the $\ce{\ce{{V}}}$-$\ce{\ce{{O}}}$ bond ($HK_{out}$).
Although these lattice configurations are un/meta-stable, they could
become stabilized under certain conditions and depending on underlying
substrates. Moreover, a combination of in- and out-of-plane buckling
can also be expected and was observed in an \emph{ab initio} calculation
of TMOs on metal substrates \citep{NogueraSubstrate}.

\begin{figure*}
\includegraphics[scale=0.63]{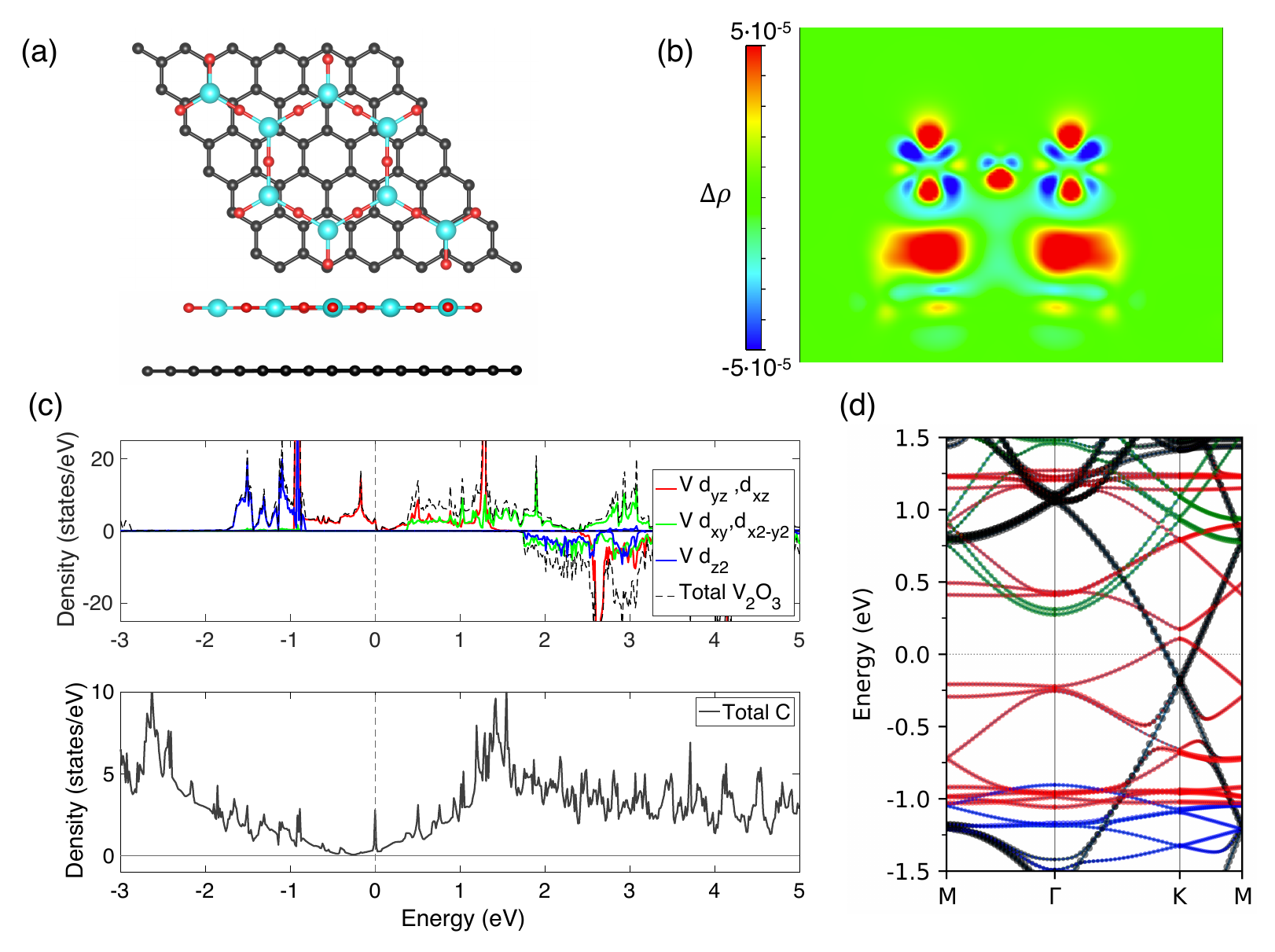}\caption{Graphene substrate study. (a) The top and side view of $(2\times2)$
$\ce{V_{2}O_{3}}$ supercell on a $(5\times5)$ graphene supercell.
(b) The calculated charge density difference $\Delta\rho$ where blue
and red correspond to excess and depleted charge density, respectively.
(c) The DOS within GGA+$U$and (d) band structure of 2D HK $\ce{V_{2}O_{3}}$
op graphene within GGA+$U$+SOC. The color code is the same used in
(c).}
\label{fig:graphene}
\end{figure*}

Identifying these different structural configurations under compressive
strain in the freestanding $\ce{V_{2}O_{3}}$ monolayer allows a further
study on the structural stability and provides a better understanding
of the involved chemical bonds governing the stability of this planar
single atomic layer. From the structural in- and out-of-plane distortions,
it is noted that the deformation of the $\ce{\ce{{V}}}$-$\ce{\ce{{O}}}$
bond develops as to prevent a further reduction of this bond length
(see Fig. \ref{fig:biaxial}c). This behavior can be linked to the
ionic character of the bond. On the other hand, from the energetic
point of view, the system prefers to undergo an in-plane deformation
rather than the out-of-plane deformation. This can be attributed to
the energy cost of the formation of a dipole moment under a buckling
of the $\ce{\ce{{V}}}$-$\ce{\ce{{O}}}$ bond.

In addition to this structural study, the electronic and magnetic
properties as a function of the biaxial strain were studied. Since
the stabilization of the metastable states ($HK$ and $HK_{out}$)
will depend on external conditions, only the most stable $HK_{in}$
configuration is considered for the compressive region. The energy
gap and the magnetic exchange interaction are shown in Fig. \ref{fig:biaxial}.
It is clear that within the strain range of $-5\%$ to $+5\%$, the
electronic and magnetic properties are sufficiently preserved to maintain
the room-temperature magnetic Chern insulating phase. It was also
confirmed that the Chern invariant $C$ remains equal to $1$, even
in the presence of the in-plane distortion of $HK_{in}$. This can
be expected as the $\ce{\ce{{V}}}$ cations still form a honeycomb
lattice, crucial for the existence of the Dirac cone \citep{existence}.
From these results, it can be concluded that the room-temperature
QAHE in 2D $\ce{V_{2}O_{3}}$ is robust against small structural deformations.

\subsection{\label{subsec:Graphene}Graphene substrate}

Motivated by the successful deposition of $\ce{Y_{2}O_{3}}$ on graphene
\citep{growthY2O3}, the feasibility of using a graphene substrate
for the growth of the $\ce{V_{2}O_{3}}$ monolayer was studied. A
$(2\times2)$ $\ce{V_{2}O_{3}}$ supercell on a $(5\times5)$ graphene
supercell was structurally optimized and that configuration is shown
in Fig. \ref{fig:graphene}a (see Supplemental Material). Similar
to $\ce{Al_{2}O_{3}}$ and $\ce{Y_{2}O_{3}}$ monolayers on graphene
\citep{HKAl2O3,HKY2O3}, the $\ce{\ce{{V}}}$ cations are located
near the top sites of the $\ce{\ce{{C}}}$ atoms to maximize the potential
bonding. The planar HK lattice structure is conserved with a minimal
distance between graphene and $\ce{V_{2}O_{3}}$ of $3.43\:\text{\AA}$,
while the $\ce{\ce{{V}}}$-$\ce{\ce{{O}}}$ bond length remains approximately
preserved.

To obtain a first estimate of the interfacial interaction strength,
the adsorbtion energy $E_{\text{{ad}}}$ is determined,

\begin{equation}
E_{\text{{ad}}}=E_{G}+E_{V_{2}O_{3}}-E_{G+V_{2}O_{3}}
\end{equation}

where $E_{G}$, $E_{V_{2}O_{3}}$ and $E_{G+V_{2}O_{3}}$ are the
total energies of the isolated graphene, isolated $\ce{V_{2}O_{3}}$,
and the combined hybrid structure, respectively. The adsorbtion energy
equals $77$ meV per $\ce{\ce{{C}}}$ atom indicating a weak interaction
between both monolayers.

\ 

To preserve the electronic properties it is important to prevent charge
transfer between both monolayers, therefore the atomic charges of
the systems were determined for the hybrid system. The average atomic
charges of both $\ce{\ce{{V}}}$ and $\ce{\ce{{O}}}$ remain unchanged,
while the average excess charge of the $\ce{\ce{{C}}}$ atom is only
$0.004e$. This indicates negligible electron transfer between both
monolayers, which can also be confirmed by the calculation of the
charge density difference $\Delta\rho=\rho_{G+V_{2}O_{3}}-\rho_{G}-\rho_{V_{2}O_{3}}$,
shown in Fig. \ref{fig:graphene}b. The charge density difference
shows a negligibly small inhomogeneous charge distribution between
both monolayers, which is mainly located between $\ce{\ce{{V}}}$
and $\ce{\ce{{C}}}$ atoms. This indicates a small orbital hybridization
of the $\pi$-bonded orbitals of $\ce{V_{2}O_{3}}$ and graphene.
This is similar to earlier studies of HK TMOs on graphene where it
was shown that vdW interactions and orbital hybridization play an
important role for the electronic properties at the interface \citep{HKAl2O3,HKY2O3}.

The electronic properties of the $\ce{V_{2}O_{3}}$ monolayer on the
graphene substrate are shown in Fig. \ref{fig:graphene}c-d. From
the band structure and DOS, it is noted that the conduction band minimum
of graphene is about $0.2$ eV below the Fermi level, indicating that
graphene is slightly n-doped. The DOS contribution of the $\ce{V_{2}O_{3}}$
remains approximately preserved, however from the calculated band
structure, it is clear that there is some hybridization of the $\pi$-bonded
orbitals. Nonetheless, the Dirac cone of $\ce{V_{2}O_{3}}$ remains
preserved with a reduction in the SOC-induced direct energy gap to
$66$ meV.

\section{Conclusions}

In this work, we have confirmed that the single atomic layer of $\ce{V_{2}O_{3}}$
with HK lattice structure is a structurally stable room-temperature
magnetic Chern insulator. This system features the coexistence of
topological order and strong electron correlations, and might therefore
be an excellent SCDS candidate. It was shown that the system can undergo
small structural deformations which preserve the honeycomb lattice
of the $\ce{\ce{{V}}}$ cations, such that the Dirac cone and the
corresponding room-temperature QAHE remain unaffected. Furthermore,
this 2D $\ce{V_{2}O_{3}}$ can be further stabilized by the support
of a graphene substrate, where there is only a small interfacial interaction
mainly attributed to orbital hybridization, but preserving the Dirac
cone of both materials. These observations together with earlier studies
on TMO with HK lattice structure \citep{HKY2O3,HKAl2O3,growthY2O3,HKMn2O3,HKCr2O3,2DTMOscreening}
should encourage the experimental investigation of this group of 2D
TMOs.

\section*{Data Availability}
The data that supports the findings of this study are provided here and the Supplementary Material.

\section*{Acknowledgements}

Part of this work was financially supported by the KU Leuven Research
Funds, Project No. KAC24/18/056 and No. C14/17/080 as well as the
Research Funds of the INTERREG-E-TEST Project (EMR113) and INTERREG-VL-NL-ETPATHFINDER
Project (0559). Part of the computational resources and services used
in this work were provided by the VSC (Flemish Supercomputer Center)
funded by the Research Foundation Flanders (FWO) and the Flemish government.

\bibliographystyle{unsrt}
\bibliography{referencesV2O3}

\clearpage

\onecolumngrid
\section*{Supplementary Material}
\subsection*{Stiffness constants}

From continuum mechanics, a general equation for the elastic energy
for a deformation of a 2D material can be obtained:

\begin{equation}
E=E_{0}+A_{0}\sum_{i}\sum_{j}\sigma_{ij}\epsilon_{ij}+\frac{A_{0}}{2}\sum_{i}\sum_{j}\epsilon_{ij}\cdot\left(\sum_{k}\sum_{l}C_{ijkl}\epsilon_{kl}\right)\label{eq:Elastic}
\end{equation}

where $E_{0}$ is the equilibrium energy, $A_{0}$ the equilibrium
unit cell area, $\sigma_{ij}$ the rank $2$ stress tensor, $\epsilon_{ij}$
the rank $2$ strain tensor, and $C_{ijkl}$ the rank $4$ stiffness
tensor. To simplify this equation with high-rank tensors, the Voigt
notation \cite{voigt1928lehrbuch} will be adopted. Furthermore, by
the symmetries of the hexagonal lattice, the stiffness matrix reduces
to a $3\times3$ matrix of the form \cite{SymmetrieElasticity}

\begin{equation}
\left(\begin{array}{ccc}
C_{11} & C_{12} & 0\\
C_{12} & C_{11} & 0\\
0 & 0 & C_{33}
\end{array}\right),
\end{equation}

with three independent stiffness constants $C_{11},$$C_{12}$ and
$C_{33}$. Since only in-plane deformations are considered, $C_{33}$
is set to zero. To determine the stiffness constants, we consider
a uni- and biaxial deformation respectively given by the strain tensors

\begin{equation}
\epsilon_{11}=\left(\begin{array}{cc}
\delta\\
0\\
0
\end{array}\right),\:\epsilon_{12}=\left(\begin{array}{cc}
\delta\\
\delta\\
0
\end{array}\right)
\end{equation}
with $\delta$ a dimensional quantity, representing the stength of
the deformation, defined by the relative change in the lattice constant
along the axis of deformation,

\begin{equation}
\delta=\frac{|a-a_{0}|}{a}
\end{equation}

where $a_{0}$ and $a$ are the lattice constants along the deformation
direction of respectively the equilibrium and deformed structure.
Using Eq. \ref{eq:Elastic}, the energy change associated with both
deformations is given by

\begin{equation}
\begin{aligned}E_{11}(\delta)=E_{0}+A_{0}\sigma_{1}\delta+\frac{A_{0}}{2}C_{11}\delta^{2}+O(\delta^{3}),\\
E_{12}(\delta)=E_{0}+A_{0}(\sigma_{1}+\sigma_{2})\delta+A_{0}(C_{11}+C_{12})\delta^{2}+O(\delta^{3}).
\end{aligned}
\label{eq:energies}
\end{equation}

\begin{figure}[h]
\centering{}\includegraphics[scale=0.4]{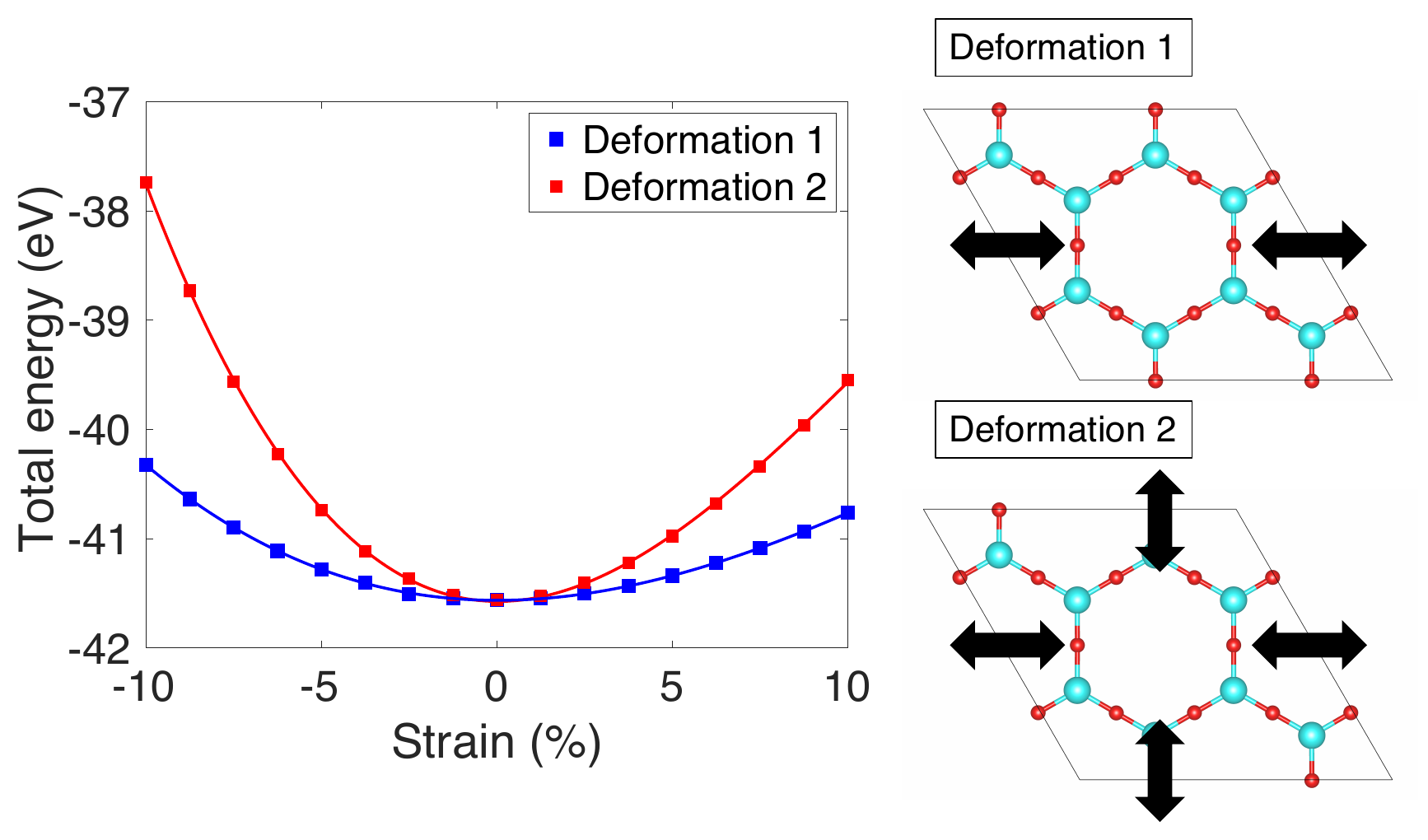}\caption{The calculated total energies for the two types of deformations shown
on the right and the corresponding fit on the left.}
\label{fig:elastic}
\end{figure}

The calculated total energies of the deformed configurations are shown
in Fig. \ref{fig:elastic} and the corresponding functions in Eq.
\ref{eq:energies} are fitted to these energies.

\subsection*{Estimate of the Hubbard correction}

This linear response calculation is based on \cite{linearresponse}
and is divided into three steps. First, a self-consistent electronic
calculation is performed without the presence of any perturbation
$\alpha$. Secondly, a non-charge-self-consistent calculation is performed
with the presence of a non-zero pertubartion $\alpha.$ In this calculation,
the charge density is not allowed to relax and thus a \emph{bare}
(non-interacting) response is obtained. Lastly, a charge-self-consistent
claculation with non-zero $\alpha$ is employed where the charge density
is allowed to relax and to screen the perturbation. This provides
the \emph{interacting} response of the system. Both bare and interacting
response can be respectively quantified by the response density functions
\cite{linearresponse}:

\begin{align}
\chi_{ij}=\frac{\partial^{2}E}{\partial\alpha_{I}\partial\alpha_{J}}=\frac{\partial n_{I}}{\partial\alpha_{J}},\\
\chi_{IJ}^{0}=\frac{\partial^{2}E^{KS}}{\partial\alpha_{I}^{KS}\partial\alpha_{J}^{KS}}=\frac{\partial n_{I}}{\partial\alpha_{J}^{KS}}
\end{align}

where $I,$ $J$ are the $\ce{\ce{{V}}}$-sites in the supercell,
and the $KS$ superscript stands for the non-interacting (Kohn-Sham)
response.

\ 

\begin{figure}[h]
\centering{}\includegraphics[width=0.8\textwidth]{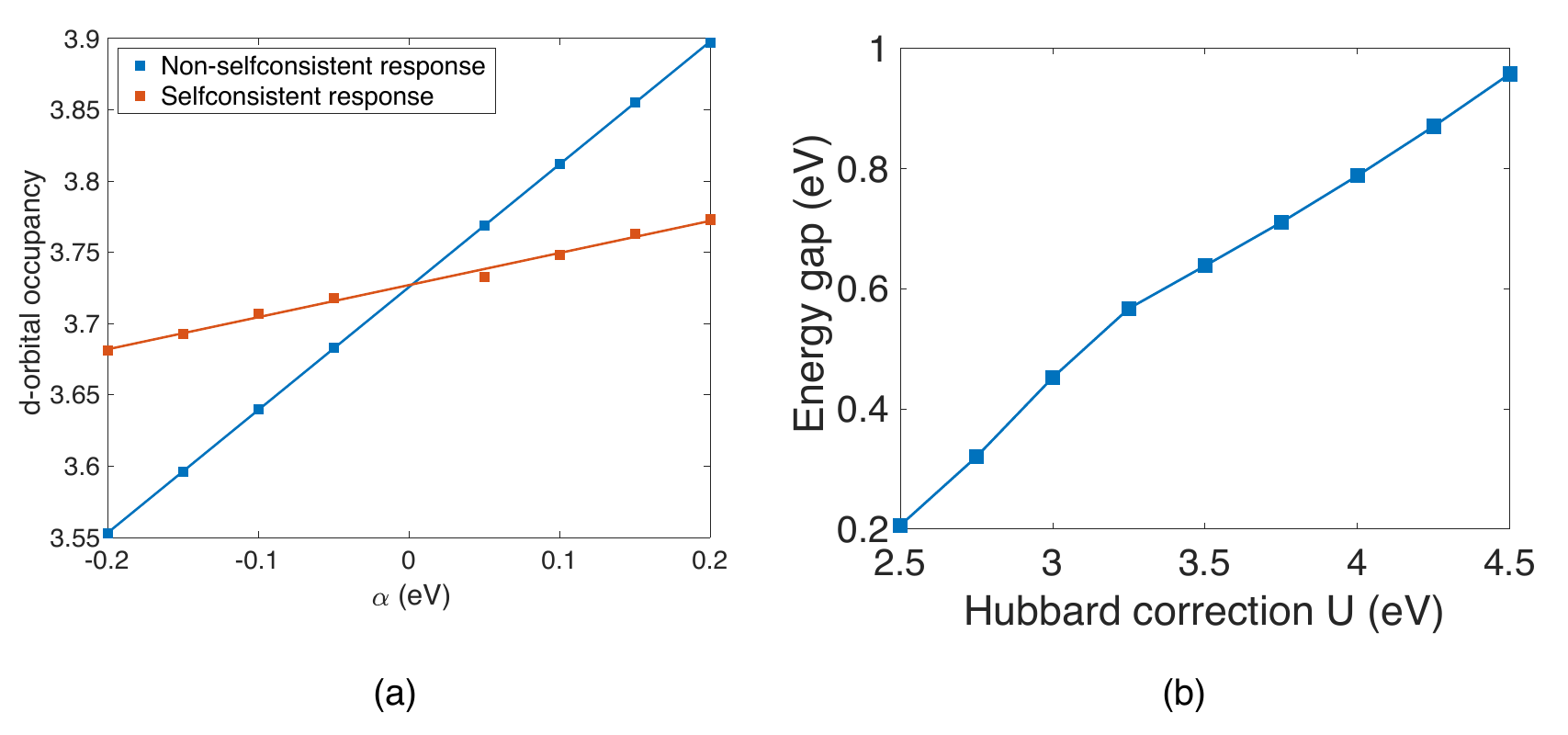}\caption{(a) The DFT calculated interacting and non-interacting response functions. (b) Indirect energy gap as a function of the Hubbard correction $U$.}
\label{fig:Hubbard}
\end{figure}

In order to obtain an accurate estimate of $U$, the calculation of
the response density functions should be repeated for increasing supercell
dimensions, until there is convergence. However, a $2\times2\times1$
supercell containing $24$ atoms should suffice for this calculation,
with larger supercells giving minor improvements. Subsequently, the
effective interaction parameter $U$ can be determined by 

\begin{equation}
U=\left(\chi_{0}^{-1}-\chi^{-1}\right)_{IJ}\label{eq:Hubbard}
\end{equation}

where $IJ$ represents the evaluation of the response density for
a perturbation a site $J$ and its response measured at site $I$.
These calculations were performed for a perturbation range of $0.4$
eV with stepsize $0.05$ eV. By linear fitting, illustrated in Fig.
\ref{fig:Hubbard}a, the response density functions were determined
from the slope and using Eq. \ref{eq:Hubbard} a Hubbard correction
of $U=3.28$ eV was found. 

\subsection*{Electronic structure in function of exchange-correlation functional}

\subsubsection*{Effect of Hubbard correction}

The electronic band structure was calculated for a Hubbard correction
ranging from $2.5$ eV to $4.5$ eV. Without the inclusion of the
spin-orbit coupling (SOC), it is observed that there is less overlap
of the different orbital levels for increasing $U$, nonetheless,
the overall effect is relatively small. By the inclusion of SOC, it
is clear that the energy gap opened at the Dirac point strongly depends
on the Hubbard correction $U$, as shown in Fig. \ref{fig:Hubbard}b.

\begin{figure}[h]
\includegraphics[scale=0.6]{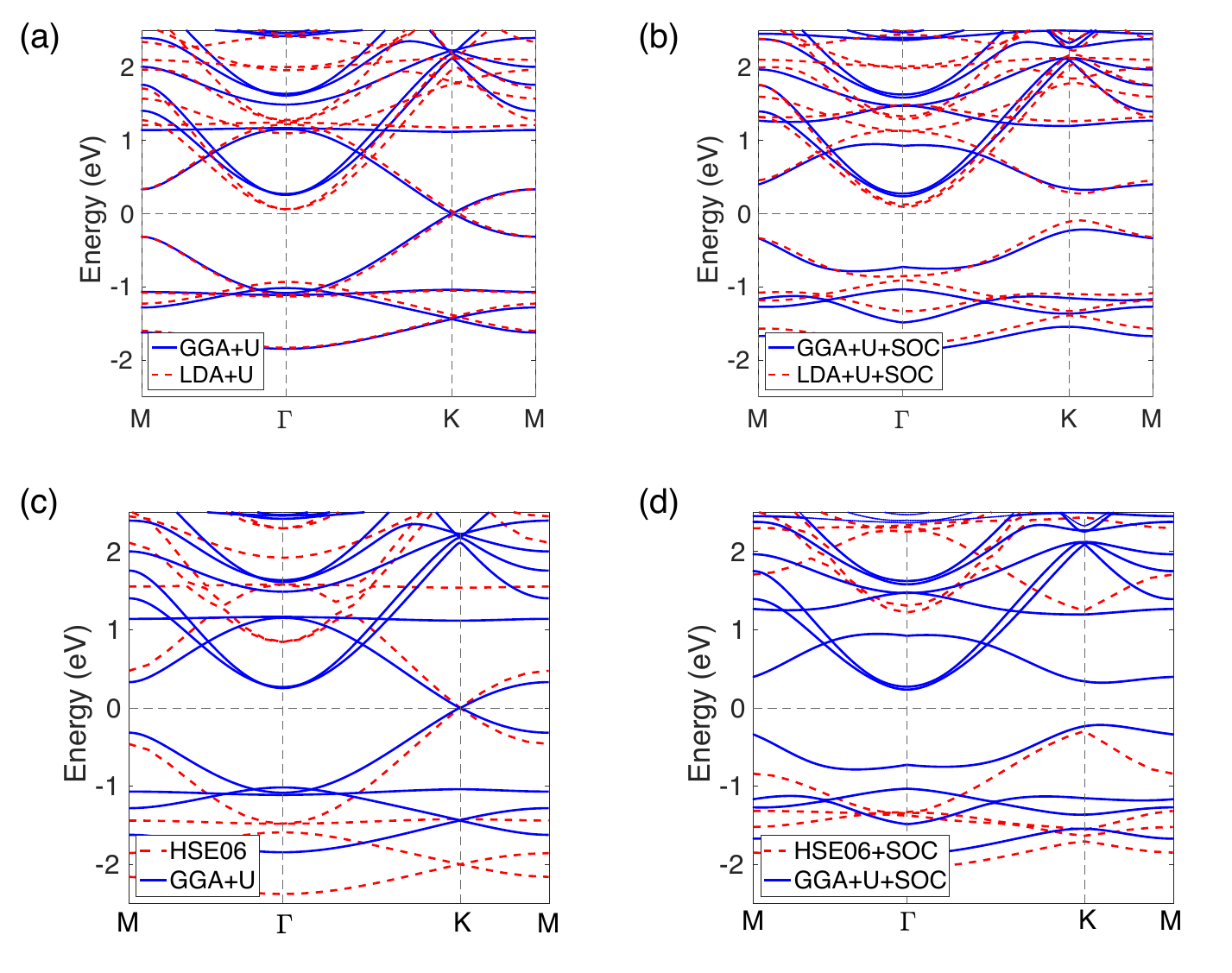}\caption{The comparison of the calculated band structure within LDA+$U$, GGA+$U$
and HSE06, with and without inclusion of SOC.}
\label{fig:bands}
\end{figure}

\subsubsection*{Exchange-correlation functional dependence}

To test the sensitvity of the results of the electronic band structure
to the choice of the functional, the three most common exchange-correlation
functionals were employed. The generalized gradient approximation
(GGA) in the form of Perdew-Burke Ernzerhof (PBE) \cite{PBE}, the
local density approximation (LDA) \cite{LDA} and the screened exchange
hybrid density functional by Heyd-Scuseria-Ernzerhof (HSE06) \cite{HSE06}.
The calculated band structures for LDA+$U$ and HSE06 are shown in
Fig. \ref{fig:bands} and compared with the band structure obtained
by GGA+$U$.

\subsection*{Topology of the band structure}

\begin{figure}[h]
\includegraphics[scale=0.3]{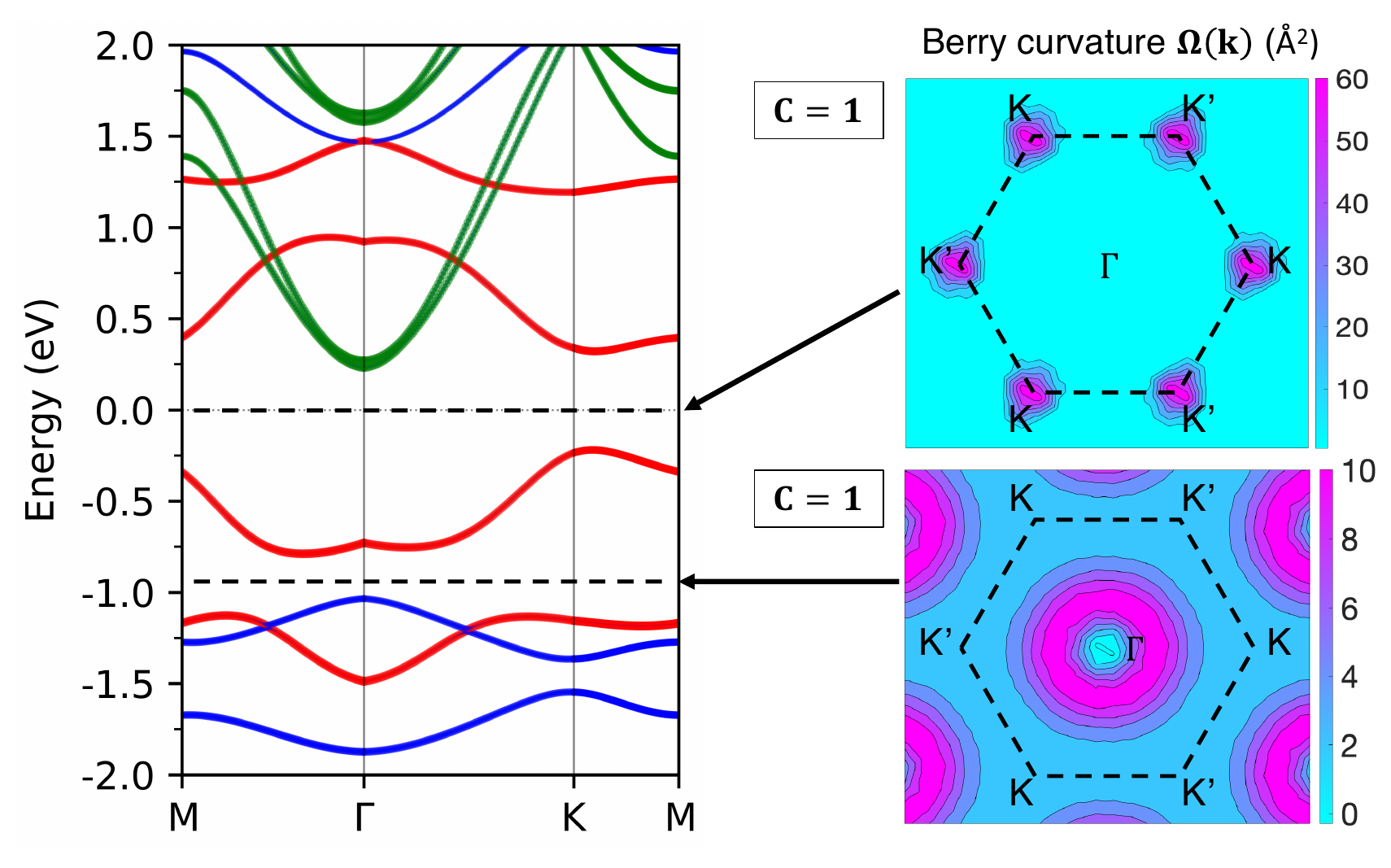}\caption{The calculated band structure and the corresponding Berry cruvature
projected onto the Brillouin zone at the Fermi level and at $-0.9$
eV.}
\label{fig:topological}
\end{figure}

In addition to the topological energy gap at the Fermi level with
a Chern number equal to one, the Chern number was also calculated
for the gap opening at the $\Gamma$ point around $-0.9$ eV below
the Fermi level. The calculated Berry cruvature is shown in Fig. \ref{fig:topological}
and was integrated by the VASPBERRY code, which is based on Fukui's
method \cite{VASPBERRY}. 

\subsection*{Graphene substrate }

\begin{figure}[h]
\includegraphics[scale=0.4]{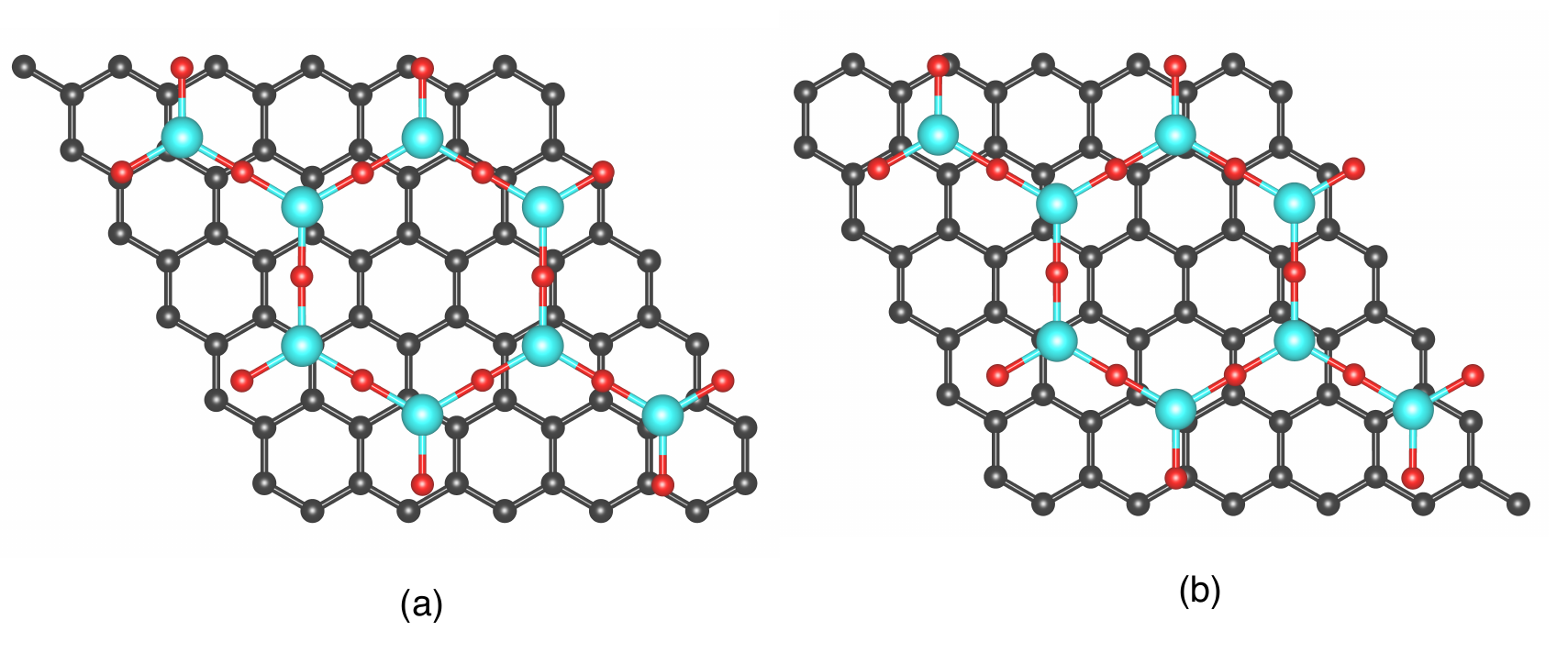}\caption{Two different superstructures with the $(2\times2)$ $\ce{V_{2}O_{3} }$
monolayer with respect to the $(5\times5)$ graphene layer. The atom
colors gray, cyan, and red corresponding to carbon, vanadium and oxygen
respectively. }
\label{fig:grapheneconfig}
\end{figure}

For the ionic relaxation of the graphene supported $\ce{V_{2}O_{3} }$
monolayer system, several configurations of the $\ce{V_{2}O_{3} }$
monolayer with respect to the graphene layer have been considered,
as shown in Fig. \ref{fig:grapheneconfig}. For all configurations,
the supercell was constructed with completely relaxed structures of
graphene and $\ce{V_{2}O_{3} }$ , with freestanding lattice constants
of $2.477\:\text{\AA}$ and $6.193\:\text{\AA}$, respectively. By
comparing the total energies of the different configurations, it was
confirmed that the other considered configurations are very close
in energy ($<10^{-3}$ eV), and as the superstructure of graphene
on $\ce{V_{2}O_{3} }$ is incommensurate it is expected that for any
configuration the $\ce{\ce{{V}}}$ cations will be located approximately
on top of the $\ce{\ce{{C}}}$ atoms. Furthermore, it was verified
that the density of states remains approximately preserved in these
other configurations. At last, the initial distance between both monolayers
was varied to confirm absolute stability. The most stable configuration
was found to have unchanged structural parameters in both monolayers,
i.e. the vdW epitaxy does not induce any significant strain within
the system.

\bibliographystyle{unsrt}
\bibliography{referencesV2O3}
\end{document}